\documentclass[11pt,letterpaper]{article}
\pdfoutput=1

\usepackage{colortbl}
 
\usepackage{graphicx,array}
\usepackage[]{xcolor}
\definecolor{darkblue}{rgb}{0,0.1,0.5}
\definecolor{darkgreen}{rgb}{0,0.5,0.2}
\definecolor{darkred}{RGB}{153,26,0}
\usepackage{ulem}
\usepackage{slashed}
\definecolor{seablue}{rgb}{0,0.2,0.6}
\usepackage{latexsym}
\usepackage{amssymb, amsmath}
\usepackage{slashed}
\usepackage{bm}        
\usepackage[numbers,sort&compress]{natbib}
\usepackage{bm,relsize}
\usepackage{slashed}
\definecolor{viola}{RGB}{134,41,198}
\usepackage{mathrsfs}
\usepackage{hyperref} 
\hypersetup{
    colorlinks=true,       
    linkcolor=darkblue,          
    citecolor=red,        
    filecolor=magenta,      
    urlcolor=blue           
}
\usepackage[all]{hypcap} 
\usepackage{subcaption}
\usepackage{booktabs}
\usepackage{array}

\usepackage[font=small,labelfont=bf,labelsep=period]{caption}

\newcommand{\be}{\begin{equation}}
\newcommand{\ee}{\end{equation}}

\newcommand{\beq}{\begin{equation}}
\newcommand{\eeq}{\end{equation}}

\begin{document}

\begin{flushright}

\end{flushright}
\vspace{.6cm}
\begin{center}
{\LARGE \bf 
BAO vs. SN evidence for evolving dark energy
}\\
\bigskip\vspace{.7cm}
{
\large Alessio Notari$^{a,b}$, Michele Redi$^c$, Andrea Tesi$^c$}
\\[7mm]
 {\it \small
$^a$ Departament de F\'isica Qu\`antica i Astrofis\'ica \& Institut de C\`iencies del Cosmos (ICCUB), \\
Universitat de Barcelona, Mart\'i i Franqu\`es 1, 08028 Barcelona, Spain\\
\vspace{.1cm}
$^b$ Galileo Galilei Institute for theoretical physics, Centro Nazionale INFN di Studi Avanzati\\
Largo Enrico Fermi 2, I-50125, Firenze, Italy\\ 
\vspace{.1cm}
$^c$INFN Sezione di Firenze, Via G. Sansone 1, I-50019 Sesto Fiorentino, Italy\\
Department of Physics and Astronomy, University of Florence, Italy
 }
\end{center}
\vspace{.2cm}

\centerline{\bf Abstract} 
\begin{quote}
We critically review the evidence for time-varying dark energy from recent Baryon Acoustic Oscillations (BAO) and Supernova (SN) observations. First, we show that such evidence is present at the 3$\sigma$ level, even without the new BAO data from the dark energy Spectroscopic Instrument (DESI), by instead using BAO data from the dark energy Survey (DES), combined with the DES5Y supernovae and Planck CMB data. Next, we examine the role of the DES5Y supernova dataset, showing that the preference for time-varying dark energy is driven by the low redshift supernovae common to both the DES5Y and Pantheon\texttt{+} compilations. We find that combining Pantheon\texttt{+} and DES5Y supernovae by removing the common supernovae leads to two different results, depending on whether they are removed from the DES5Y or the Pantheon\texttt{+} catalog, leading to stronger or weaker exclusion of $\Lambda$CDM, at the (3.8$\sigma$) and (2.5$\sigma$) level, respectively. These common supernovae have smaller error bars in DES5Y compared to Pantheon\texttt{+}, and, as recently pointed out, there is an offset in magnitude in DES5Y between supernovae at ($z > 0.1$), where almost all the measurements taken during the full five years of DES are, and the low-redshift ones ($z < 0.1$), where all the historical set of nearby supernovae lies. We show that marginalizing over such an offset in DES5Y would lead to significantly weaker evidence for evolving dark energy.
\end{quote}

\vfill
\noindent\line(1,0){188}
{\scriptsize{ \\ E-mail:\texttt{   \href{notari@fqa.ub.edu}{notari@fqa.ub.edu}, \href{mailto:michele.redi@fi.infn.it}{michele.redi@fi.infn.it}, \href{andrea.tesi@fi.infn.it}{andrea.tesi@fi.infn.it}}}}
\newpage

\section{Introduction and status}
Learning that dark energy (DE) changes in time would be the greatest scientific result since the discovery of the accelerated expansion of the Universe itself, and it would have implications for fundamental physics both at the phenomenological and theoretical level. For this reason testing if DE is just a cosmological constant ($\Lambda$) or if it possibly has a time evolution is a very active experimental and observational field as it will establish the range of validity of the $\Lambda$CDM model.

Since the Universe has just ``recently'' entered its accelerated expansion phase, the search for time dependencies of dark energy has to rely on observables that are sensitive to the scale factor, $a(t)$, of the Friedmann-Robertson-Walker (FRW) metric background, at relatively recent cosmic times $t$, or redshifts $z$, on the possible largest scales.

It turns out that a very useful quantity is the so-called transverse diameter distance $D_M(z)$, defined -- for a spatially flat universe -- as the integral of the inverse Hubble parameter
\be
D_M(z)\equiv  \int_0^z \frac{dz'}{H(z')}\,.
\ee
The Hubble scale $H(z)$ is determined by the total energy density of the Universe at each redshift, through the Friedmann equation, and therefore $D_M$ is sensitive to a possible time evolution of DE, which at late time dominates the budget, $\Omega_{\rm DE}\approx 70\% $. Luckily, such a quantity $D_M$ is directly related to two of the main observables available in the late universe: the luminosity flux from supernovae (SN), and the angle related to the so-called Baryon Acoustic Oscillations (BAO). It comes with no surprise that these two have received great attention in recent times.

In particular, the recent determination of the BAO at different redshift bins by the dark energy Spectroscopic Instrument (DESI)~\cite{DESI:2024mwx} has sparked an intense debate on the nature of DE (see for example \cite{DESI:2024aqx,DESI:2024kob}). The claim can be quickly summarized by saying that the collaboration found significant evidence for an evolving DE component.

This conclusion has been reached thanks to two main steps. First, the collaboration has tested the hypothesis that DE is parametrized as a fluid with a time-dependent equation of state $w$,
\be\label{eq:CPL}
w(a)=w_0 + w_a \frac {z}{1+z}\,,
\ee
following the Chevallier-Polarski-Linder parametrization \cite{Chevallier:2000qy,Linder:2002et}. Second, the final result has been reached -- crucially -- by combining the DESI BAO measurements with Planck CMB data in combination with a SN dataset.  This leads to $w_0w_a$CDM favoured compared to  $\Lambda$CDM at the $2.5\sigma, 3.5\sigma$ and $3.9\sigma$ level depending on the SN dataset considered, i.e. Pantheon\texttt{+}~\cite{Pantheon+,likePP}, DES5Y \cite{DES:2024jxu} or Union3~\cite{union3} SN datasets respectively.

Both steps can be reconsidered in a more complete analysis.

However, despite the fact that the best-fit values for the new parameters $w_0$ and $w_a$ correspond to a fluid with a pathological equation of state at early times ($w<-1$), we will not reconsider here the theoretical model, and we will use the same parametrization of \eqref{eq:CPL}. Let us notice, however, that consistent models (which can be realized with a simple and healthy quintessence scalar field) in agreement with the claim of \cite{DESI:2024mwx} have already been presented in the literature \cite{Notari:2024rti,Ramadan:2024kmn, Bhattacharya:2024hep,Andriot:2024jsh,Tada:2024znt,Gialamas:2024lyw,Giare:2024gpk}. We refer to those papers for all the theoretical considerations, and we will leave the discussion aside in this work.

Instead, the primary focus of our work is to carefully re-examine the claims in~\cite{DESI:2024mwx} from the point of view of data analysis, taking inspiration by the already known fact that, when the combination of DESI BAO is done with different SN datasets, the evidence for evolving DE diminishes. For example, while the evidence is sizable using DES5Y or the Union3 datasets, milder evidence comes from using the Pantheon\texttt{+} \cite{Brout:2022vxf} SN dataset. It is of primary importance to understand whether the evidence is driven by one particular dataset, or if it supported by several of them. We will review thus the role of BAO itself and of SN datasets, to better understand how solid is the evidence coming from each of them.

The aim of our data analysis is actually two-fold. 

First, we will show that the DESI BAO data can be replaced by another recent measurement, namely the year-6 BAO measurement from the dark energy Survey (DES)~\cite{DES:2024pwq}, which consists of a single bin with $2.1\%$ precision at $z=0.85$. Combining this measurement with Planck and DES5Y SN we will compare to the fit that includes the DESI BAO data, with respect to the preference for evolving 
dark energy against $\Lambda$CDM.

Second, we will analyze the difference between SN datasets used in the analysis. In particular we look for evidence of evolving DE when combining together DES5Y with Pantheon\texttt{+}, by removing the common dataset. We will show, however, that this procedure is not unique, due to discrepancies in such a common dataset between the two catalogs.  Indeed it has been recently pointed out~\cite{Efstathiou:2024xcq} that in the DES5Y catalog the common SN have a rather different trend as a function of $z$, compared to the same SN in Pantheon\texttt{+}. In particular it has been claimed that such a common dataset in DES5Y has a $\sim$0.04 offset in magnitude between the low and high-redshift SN, which was not present in Pantheon\texttt{+}, and that once this offset is removed DES5Y leads to SN fits which are fully consistent with Pantheon\texttt{+}. For this reason we further explore the role of the common SN dataset  in a more general analysis,  leaving a free relative offset in magnitude  between low and high redshift SN, and fitting to the full CMB+BAO+SN dataset.

The paper is organized as follows. In the next section we summarize the observables and the datasets used in the analysis. In section \ref{sec:BAO} we show the results by varying different BAO datasets. Later, in section \ref{sec:SN} we fix the BAO from DESI and explore the impact of different SN catalogs, by carefully treating SN common to both DES5Y and Pantheon\texttt{+} datasets. We summarize our findings in section \ref{sec:conclusions}. In appendix \ref{app:tables} we report the  best fit parameters, mean and confidence intervals, based on the $w_0w_a$CDM parametrization determined through the Markov-Chain-Monte-Carlo for different datasets. Technical details of the SN datasets can be found  in appendix \ref{app:binning}.

\begin{table}[t!]
\centering
\renewcommand{\arraystretch}{1.1} 

\setlength{\tabcolsep}{5pt}  

\begin{tabular}{>{\columncolor[gray]{0.9}}l | >{\columncolor[gray]{0.97}}l | >{\columncolor[gray]{0.97}}p{10cm} | >{\columncolor[gray]{0.97}}l}

\rowcolor[HTML]{3D3E50} 
\textcolor{white}{\textbf{Category}} & \textcolor{white}{\textbf{Name}} & \textcolor{white}{\textbf{Description}} & \textcolor{white}{\textbf{Ref.}} \\

\rowcolor[gray]{0.92} 
\textbf{CMB} & P18 & {\small Planck 2018 high-$\ell$ TT, TE, EE ; low-$\ell$ TT ; low-$\ell$ EE likelihoods; Planck 2018 lensing data.} & \cite{Planck:2019nip} \\
\hline

\rowcolor[gray]{0.97} 
\textbf{SN} & Pantheon\texttt{+} &{\small Pantheon\texttt{+} supernovae compilation.} & \cite{Pantheon+,likePP} \\ 

\cline{2-4}
\rowcolor[gray]{0.97} 
& DES5Y & {\small DES5Y supernovae compilation. The likelihood has been derived by us using data and covariance found in \cite{DES-data}, and it has been already used in \cite{Notari:2024rti}  and cross-checked against the likelihood given in~\cite{DEScobaya}. }& \cite{DES:2024tys}.\\

\hline
\rowcolor[gray]{1} 
\textbf{BAO} & DESI$_{\rm BAO}$ & {\small DESI 2024 BAO measurements. Redshift bins: $z = [0.3, 0.51, 0.71, 0.93, 1.32, 1.49, 2.33]$} & \cite{DESI:2024mwx} \\
\cline{2-4}

\rowcolor[gray]{1} 
& DES$_{\rm BAO}$ & {\small DES BAO measurement at effective redshift $z = 0.85$} & \cite{DES:2024pwq} \\

\cline{2-4}
\rowcolor[gray]{1} 
& BOSS$_{\rm BAO}$ & 
{\small BAO measurements from 6dFGS at $z = 0.106$ \cite{Beutler:2011hx}, SDSS MGS at $z = 0.15$ \cite{Ross:2014qpa} (BAO smallz), and CMASS and LOWZ galaxy samples of BOSS DR12 at $z = [0.38, 0.51, 0.61]$ \cite{BOSS:2016wmc}}& \cite{Beutler:2011hx,Ross:2014qpa,BOSS:2016wmc}

\end{tabular}
\caption{\label{tab:all-data}Datasets used in our work divided by categories (CMB, SN, BAO).}
\end{table}

\section{Observables and data}\label{sec:observables}
Our analysis depends on the cosmological model under consideration. As stated in the introductory section, we only consider two models: $\Lambda$CDM vs $w_0 w_a$CDM. The latter refers to a DE component with equation of state phenomenologically parametrized in~\eqref{eq:CPL}, which implies that its energy density evolves as,
\be\label{eq:CPL_budget}
\rho_{\rm DE}= \rho_c \Omega_{\rm DE} \, \exp (3 w_a(a-1)) a^{-3(1+w_0+w_a)}\,.
\ee
Here $\rho_c\equiv 3 M_{\rm Pl}^2 H_0^2$ is the critical density and we assume a spatially flat universe so that the energy densities $\rho_i$ of DE, matter and radiation satisfy $\Omega_{\rm DE}+ \Omega_m+ \Omega_r=1$, where $\Omega_i\equiv\rho_i/\rho_c$. Notice that for $w_0=-1$ and $w_a=0$ one recovers the cosmological constant. 

CMB observables (meaning TT, TE, EE spectra and lensing) are important to the discussion because they constrain the input cosmological parameters, and they are always included in our analysis. Also, although the fluctuations of the $w_0w_a$CDM component have a limited impact on the CMB (see however \cite{Notari:2024rti} for more details), we always include them in our numerical computations following the prescription of \cite{Fang:2008sn} when the equation of state crosses the `phantom' divide, i.e. $w<-1$.

Of more direct connection with the focus of the paper are the observables of the BAO scales and SN luminosity fluxes.

\paragraph*{BAO}~\\
The BAO ruler is the sound horizon at the epoch of baryon drag $r_d$, at redshift $z_d$, which is computed by an integral
\be
r_d\equiv \int_{z_d}^\infty dz' \frac{c_s(z')}{H(z')} \, ,
\ee
where $c_s(z)$ is the sound speed of the photon-baryon fluid. The measurements of the BAO scale are reported as  dimensionless ratios, i.e. the angular scale $D_M(z)/r_d$ for the transverse direction, although different quantities and combinations are also used \cite{DESI:2024mwx}.

\paragraph*{SN fluxes}~\\
Calibration of SN fluxes can be more subtle. The observed flux $F$ of a Supernova with intrinsic luminosity $L$ is a function of $D_M$, since $F\equiv L/(4\pi d_L^2)$, where the luminosity distance turns out to be $d_L=(1+z)D_M(z)$, i.e. the ``distance duality"  relation~\cite{Bassett:2003vu,EUCLID:2020syl}. It is common to report the magnitude:
\be\label{eq:mu}
\mu(z)\equiv 5 \log_{10}[(1+z)D_M(z)] - M \,.
\ee
For our analysis here $M$ is simply a nuisance parameter~\footnote{We note that Pantheon\texttt{+} and DES5Y catalogs have different conventions, and their $M$ has a relative offset $\Delta M \approx 19.35$.}, since it  depends both on $L$ and on the present-day Hubble rate $H_0$~\footnote{Fixing this by the ``distance ladder" method, as done by~\cite{Riess:2021jrx} would lead to a severe ``Hubble tension" in both $\Lambda$CDM and $w_0w_a$CDM models. See instead~\cite{Allali:2024cji} for a recent analysis with DESI data that can address the Hubble tension, by considering dark radiation models that modify $r_d$ in the early Universe.}.
The SN datasets report the mean value and error of $\mu$ per each observation and their correlations via the covariance matrix.

\subsection{Datasets used in this work}
All the above observables are computed with the \texttt{CLASS}~\cite{CLASS-II}  Boltzmann solver. Our aim is to perform a Bayesian analysis to infer the posterior distributions of our cosmological parameters, and the Markov Chain Monte Carlo (MCMC) samples are generated through \texttt{MontePython}~\cite{Audren:2012wb, Brinckmann:2018cvx}. The MCMC samples are then analyzed with \texttt{GetDist}~\cite{getdist} to produce all the posterior distributions and all the triangle plots of our paper.

In this work we consider three classes of datasets, corresponding to CMB, BAO and SN samples, as reported in table \ref{tab:all-data}, to which we refer for the names used in the plots and the description of all individual datasets and likelihoods.

\section{BAO likelihoods: DES vs DESI}\label{sec:BAO}

\begin{figure}[t]
  \centering
  \includegraphics[width=.95\textwidth]{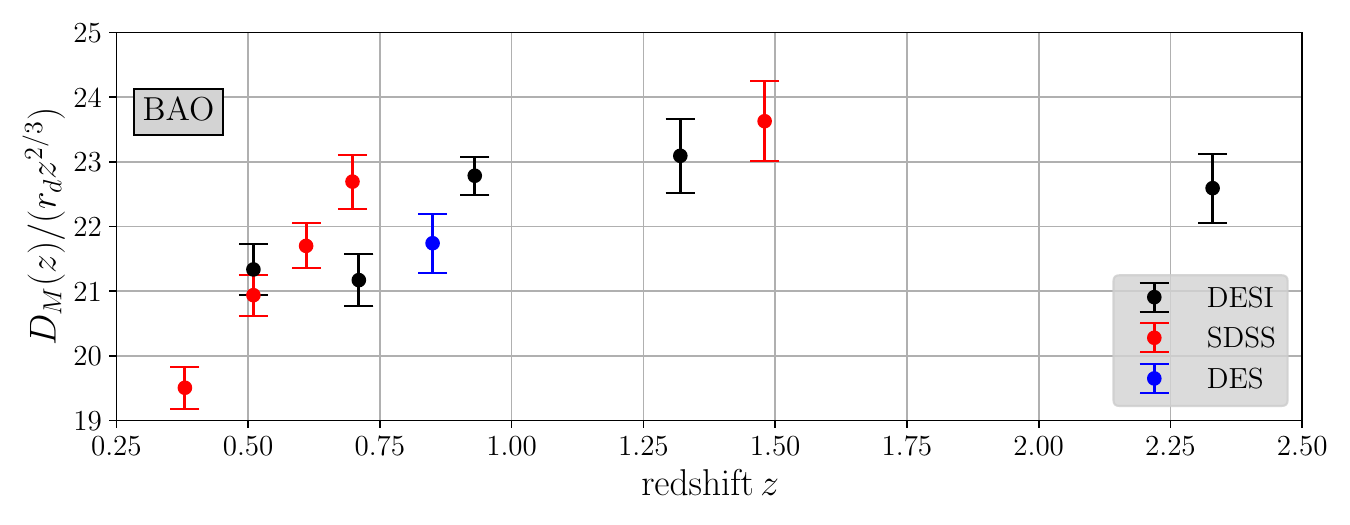}
       \caption{\it Part of the BAO dataset used in our work. For convenience we show $D_{M}(z)/(r_d z^{2/3})$.}
        \label{fig:BAO}
\end{figure}

In this section we are going to analyze the $w_0 w_a$CDM model with different  combinations of datasets. In particular we are after the impact of the recent DESI and DES measurements of BAO scale. 
As discussed in the introduction, the recent DESI measurements of the BAO have attracted a lot of attention as they show a preference for evolving dark energy with equation of state $w\ne -1$. 
This happens when DESI is combined with CMB and DES5Y supernovae dataset. We refer the reader to \cite{DESI:2024mwx} for all the details of the experimental analysis.

Here we combine a different BAO dataset to establish whether the evidence against $\Lambda$CDM found in \cite{DESI:2024mwx} is confirmed at similar levels. 
The focus is to compare two different dataset combinations that only differ by the BAO set used, namely
\be\label{BAO2}
\begin{split}
&1)\quad \mathrm{P18+DES5Y+BAO_{\rm DESI}}\,,\\
&2)\quad \mathrm{P18+DES5Y+BAO_{\rm DES}}\,.
\end{split}
\ee
whose likelihoods are listed in table \ref{tab:all-data}. We also compare with previous BAO measurements from BOSS. For a visualization of part of the dataset see figure \ref{fig:BAO}.

\begin{figure}[h!]
  \centering
  \includegraphics[width=0.79\textwidth]{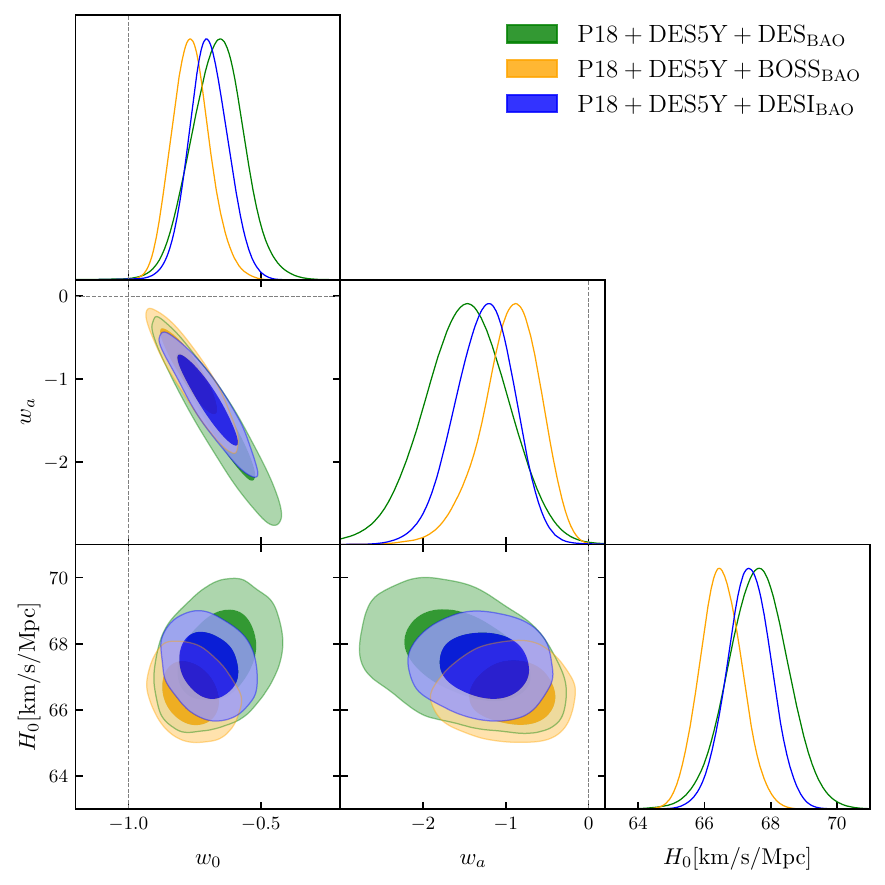}~
\caption{\it \small Marginalized posterior distributions for the parameters $w_0$, $w_a$ and $H_0$ in the $w_aw_0$CDM model. All input parameters have uniform priors. The datasets included are P18 + DES5Y and BAO dataset (DES, in green; DESI, in blue; BOSS, in orange), as defined in table \ref{tab:all-data}. }
      \label{fig:DESBAO1}
\end{figure}

The measurement of the BAO scale in Ref.~\cite{DES:2024pwq} from the DES collaboration, that appeared before DESI, consists of a single very precise data point at $z=0.85$ and it also shows a discrepancy with the $\Lambda$CDM best fit to the Planck data at  $2.1\sigma$. It is thus interesting to consider the impact of this result on the global fit when other 
measurements are also included. We thus perform a Bayesian analysis of evolving dark energy, replacing the DESI$_{\rm BAO}$ with the DES$_{\rm BAO}$ datapoint in the analysis of Ref.  \cite{DESI:2024mwx}.

We show in fig.~\ref{fig:DESBAO1} and Table~\ref{tab:w0waDESBAO} that DESY5 supernovae + DES$_{\rm BAO}$ (combined with Planck 2018 CMB TT+TE+EE spectra and Planck 2018 lensing data, P18 in short) has evidence against $\Lambda$CDM. With such a dataset indeed the best-fit of $w_0w_a$CDM has a $\Delta\chi^2=-12$ compared to $\Lambda$CDM, leading to an exclusion of the latter model with a confidence level of $99.75\%$, i.e. $3 \sigma$, in the same direction of the $3.9\sigma$ exclusion obtained when using DESI as a BAO dataset. We have also performed a similar fit adding the DES$_{\rm BAO}$ bin to the DESI$_{\rm BAO}$ dataset, finding negligible differences with respect to the case where the BAO dataset is taken from DESI alone, which shows that there is good agreement between these two BAO datasets,  providing a consistent picture (in contrast with previous BOSS data, which lead instead to quite different preferred regions: closer to $\Lambda$CDM and with lower $H_0$, see fig.~\ref{fig:DESBAO1}).

\section{Supernovae likelihoods: DES5Y vs Pantheon\texttt{+}}\label{sec:SN}
In this section we critically review the role of supernovae in providing evidence for evolving DE.
Our aim is to assess the role of different SN datasets, focusing on DES5Y and Pantheon\texttt{+} and their possible combinations (see table \ref{tab:all-data} for all details).~\footnote{A similar analysis was not possible for the Union3 dataset \cite{union3}, which is not publicly available in unbinned form.}

The DES5Y catalog has 1829 SN, including 335 supernovae in common with Pantheon\texttt{+}.  The latter catalog has 1701 entries, but there are actually only 1543 supernovae (several SN appear more than once, since they actually are the same SN observed in different surveys). There are 375 entries in common with DES5Y (which correspond to 335 individual supernovae). 

A preliminary summary of the composition of the two catalogs is provided in Table~\ref{tab:mytable}. We also show the sample resulting in removing from DES5Y the supernovae in common with Pantheon\texttt{+} (and viceversa). We call such samples, defined by having no supernovae in common between the two,  $ \overline{\rm DES5Y}$  and  $ \overline{\rm Pantheon\texttt{+}}$ respectively.

\begin{table}[h]
    \centering
    \renewcommand{\arraystretch}{1.25}  
    \begin{tabular}{c||c||c|c||c||l|l}
        Catalog & \# tot. & {\small$\#  (z<0.1)$} & {\small$\#  (z>0.1)$} & \# SNe & \# tot. (com.) &  \# SNe (com.)  \\ \hline\hline
        Pantheon\texttt{+}  & 1701   & 741 &  960 & 1543 & 375 {\tiny (w/DES5Y)}   & 335 {\tiny (w/DES5Y)} \\
        DES5Y  & 1829   & 197 & 1632 & 1829 & 335 {\tiny (w/Pantheon\texttt{+})}  & 335 {\tiny (w/Pantheon\texttt{+})}  \\ \hline
        $ \overline{\rm Pantheon\texttt{+}}$   & 1326   & 511  & 815 & 1208 & 0 {\tiny (w/DES5Y)}   & 0 {\tiny (w/DES5Y)} \\
        $ \overline{\rm DES5Y}$  & 1494   & 7 & 1487 & 1494 & 0 {\tiny (w/Pantheon\texttt{+})}  & 0 {\tiny (w/Pantheon\texttt{+})}  \\ \hline
    \end{tabular}
    \caption{\it Supernovae catalogs used (and constructed) in this work and their compositions. The sample $\overline{\rm Pantheon\texttt{+}}$ and $\overline{\rm DES5Y}$ are obtained by removing the SN in common with DES5Y and Pantheon\texttt{+} respectively.}
    \label{tab:mytable}
\end{table}

\subsection{Combining the Pantheon\texttt{+} and DES5Y supernovae datasets}\label{sec:COMBO}

\begin{figure}[th]
  \centering
  \includegraphics[width=0.76\textwidth]{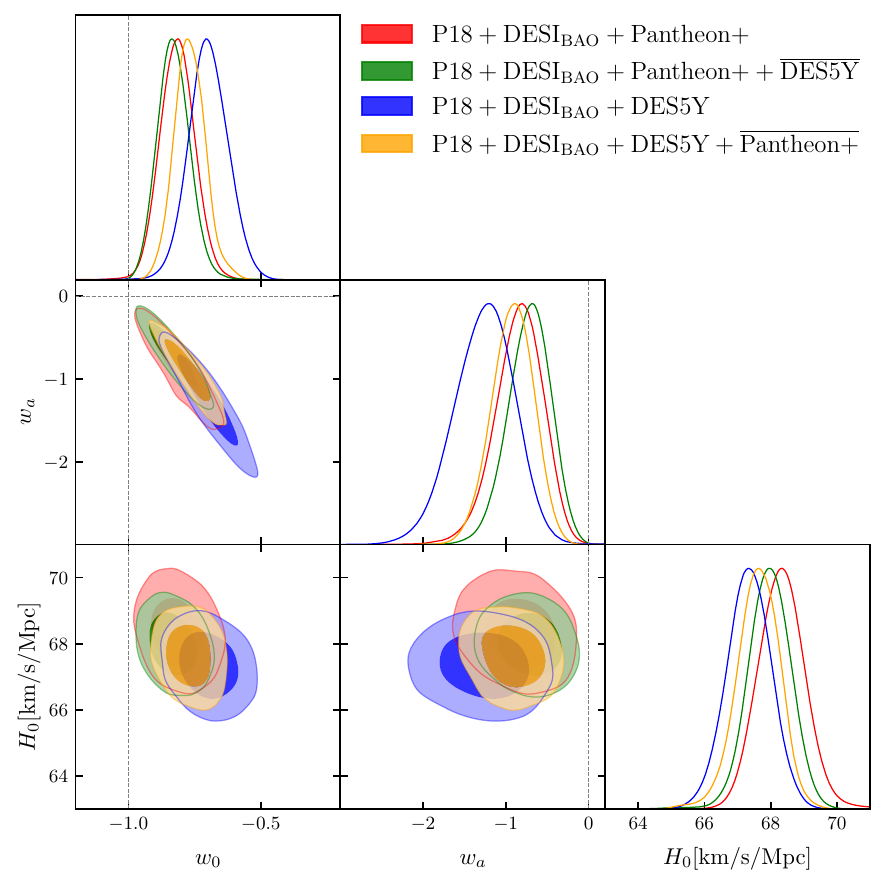}~
\caption{\it \small Marginalized posterior distributions for the parameters $w_0$, $w_a$ and $H_0$ in the $w_aw_0$CDM model. All input parameters have uniform priors. The datasets included are P18 + DESI-BAO and a Supernovae sample. In particular we have tested four choices of SNe samples (DES5Y, blue; Pantheon\texttt{+}, red;  Pantheon\texttt{+} + $\overline{\text{DES5Y}}$, green; DES5Y + $\overline{\text{Pantheon\texttt{+}}}$, orange). Definitions are found in tables \ref{tab:all-data} and \ref{tab:mytable} and in section \ref{sec:COMBO}.} 
      \label{fig:DESPANcombo}
\end{figure}

Given the relevance of supernovae to establish the evidence for evolving dark energy it is tempting to try to include all the available data
to increase the statistical significance. However it turns out that, due to a different treatment of the supernovae in common, such a combination is  not univocal. 
Crucially this can be achieved in two ways: by removing the common dataset from  DES5Y (obtaining the $\overline{\rm DES5Y}$ dataset) or from Pantheon\texttt{+} (obtaining the $\overline{\rm Pantheon\texttt{+}}$ dataset), see table \ref{tab:mytable} for details. Once the common set is removed from one catalog we then combine with the other. 
While this procedure does not eliminate all the correlations between the datasets we believe that this first-stage analysis will be useful to determine the constraining power of the precise combination and the consistency of the two datasets.

In order to show the impact of these preliminary combinations we also combine Planck and DESI BAO datasets together with different SN samples. 
In particular we use the following combinations:
\be\label{combo}
\begin{split}
&1)\quad \mathrm{P18+DESI+DES5Y}\,,\\
&2)\quad \mathrm{P18+DESI+Pantheon\texttt{+}}\,,\\
&3)\quad \mathrm{P18+DESI+\overline{{\rm DES5Y }} +  {\rm Pantheon\texttt{+}}}\,,\\
&4)\quad \mathrm{P18+DESI+\overline{{\rm Pantheon\texttt{+} }} +  {\rm DES5Y}}\,.
\end{split}
\ee
We perform the statistical analysis for $\Lambda$CDM and $w_0w_a$CDM in all four cases. The first two cases are again validated against the analysis of~\cite{DESI:2024mwx}.

We find different results depending on which set of supernovae is discarded.
The combination 3) where DESY5 common supernovae are discarded leads to results that are close to the case 2) where only the Pantheon\texttt{+} catalog is included. This case is in milder disagreement with $\Lambda$CDM, i.e. disfavored at $\sim 2.5\sigma$. This is also consistent with the statement of Ref.~\cite{Efstathiou:2024xcq}, as discussed in the next subsection, that the discrepancy with $\Lambda$CDM is mostly driven by the common subset in DES5Y. For the combination 4), where Pantheon\texttt{+} common SNe are discarded we get instead a much stronger exclusion of $\Lambda$CDM, at about $3.8\sigma$, similar to the case of DES5Y alone.  

The fact that different  combinations, that only differ for common SNe, lead to quantitatively different conclusions  urges the need for a deeper understanding of the difference between Pantheon\texttt{+} and DESY5 datasets to which we now turn.

\begin{figure}[t]
  \centering
  \includegraphics[width=1.05\textwidth]{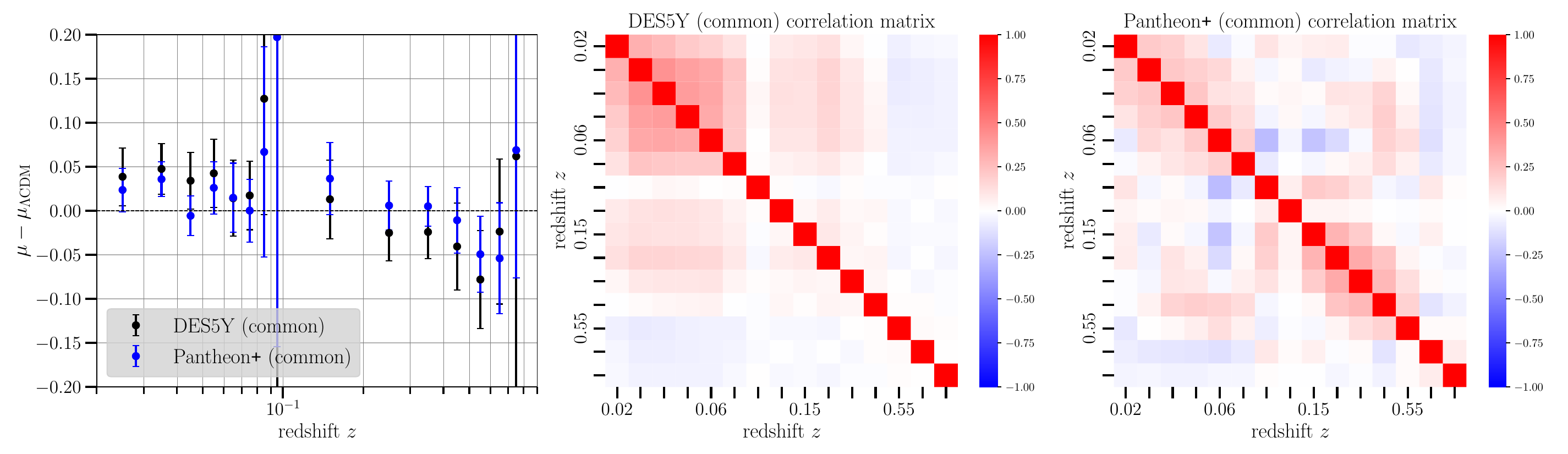}
       \caption{\it Pantheon\texttt{+} and DES5Y common (binned) data samples.  The weighted binning has been done as discussed in appendix \ref{app:binning}. In the left panel we show the weighted binning of SN magnitudes with respect to a $\Lambda$CDM reference model  (see text) for both DES5Y and Pantheon\texttt{+} for the SN in common between the two sample. In the center and right panels we show the binned correlation matrix of the corresponding samples.}
        \label{fig:PAN_DES_common}
\end{figure}

\subsection{Splitting the DES5Y supernovae dataset}\label{sec:SPLIT}

\begin{figure}[t]
    \centering
    \begin{minipage}{0.45\textwidth}
        \includegraphics[width=\linewidth]{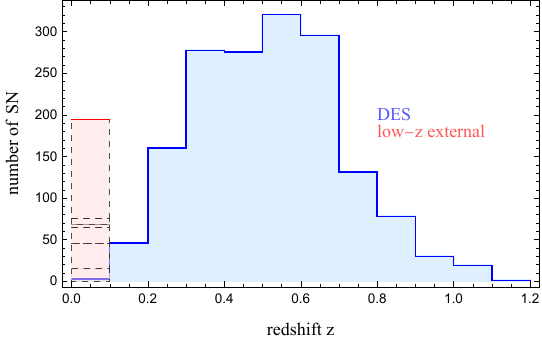}
        \caption*{}
    \end{minipage}\hfill
    \begin{minipage}{0.45\textwidth}
        \captionsetup{type=table} 
        \centering
{\small        \begin{tabular}{r|c|c}
            Surveys (ID) & $\#\mathrm{SN}\ (z<0.1)$ &  $\#\mathrm{SN}\ (z>0.1)$  \\ \hline
            DES \cite{DES:2024jxu} & 3   & 1632   \\ 
            CFA3S \cite{CfA3} & 15 & 0 \\
            CFA3K\cite{CfA3} & 31& 0\\
            CFA4p2\cite{CfA4} & 19 &0 \\
            CFA4p3 \cite{CfA4}& 3 & 0\\
            CSP \cite{CSP} & 8 & 0\\ 
            FOUND \cite{FOUND}& 118 &0\\\hline
            total & 197 & 1632
                 \end{tabular}}
                   \caption*{}
 \end{minipage}
   \includegraphics[width=.97\textwidth]{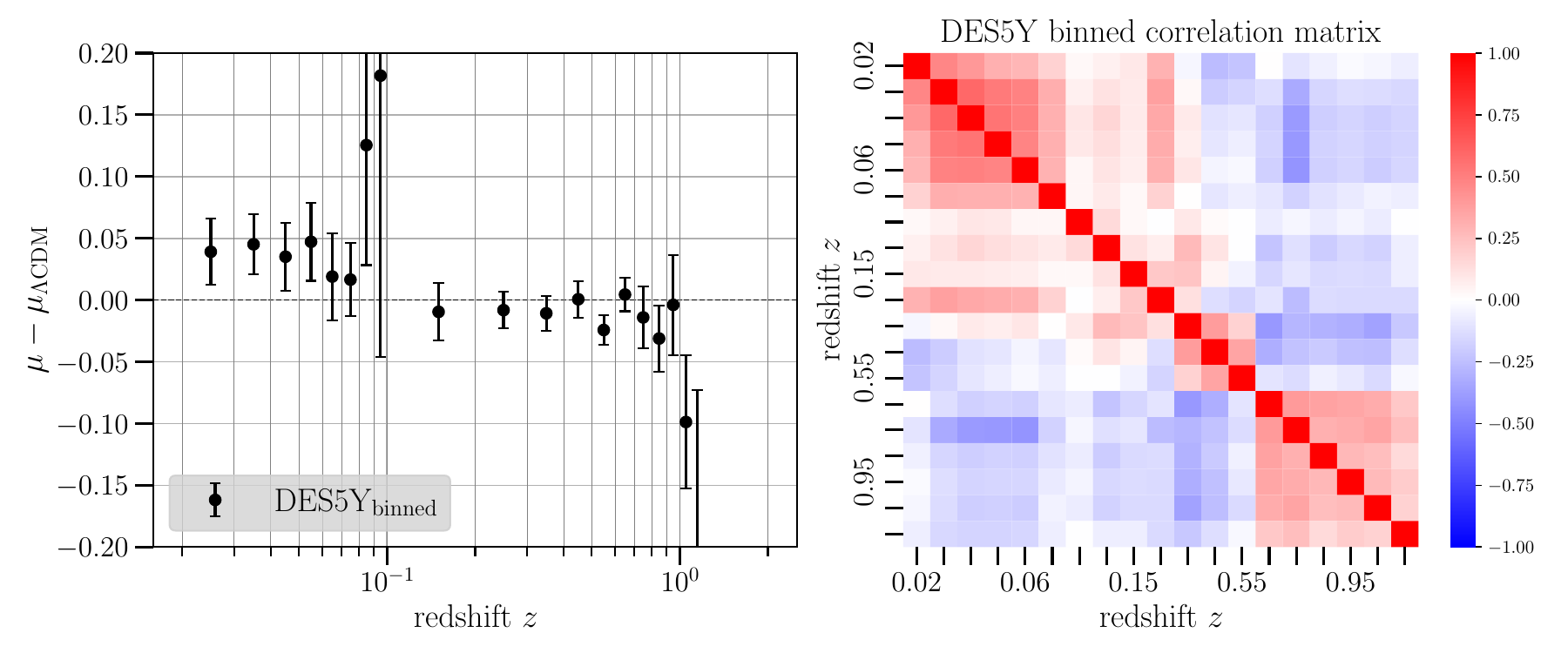}
    \caption{\it  Composition of the full DES5Y sample. Upper Left:  number of SN in different redshift bins of the DES5Y catalog. The low-z external data are shown in a stacked histogram. Upper Right: decomposition of DES5Y by surveys (at low and high redshift) as found on the dataset available from \cite{DES-data}.  Lower left: weighted binning of SN magnitudes with respect to a $\Lambda$CDM reference model (see text).  Lower right: Binned correlation matrix of the DES5Y sample. }
    \label{fig:composition}
\end{figure}

The inspection of the common sample between DES5Y and Pantheon\texttt{+} can have other implications. Indeed, recently, it has been pointed out in Ref.~\cite{Efstathiou:2024xcq} that the supernovae in common between Pantheon\texttt{+} and DES5Y appear, in the DES5Y catalog, with an offset  between low ($0.025<z<0.1$) and high ($z>0.1$) redshift magnitudes of about 0.04, due to different bias corrections applied in the DES5Y catalog.  Let us comment on the fact that the composition of the DES5Y catalog strongly differs between low-$z$  and high-$z$ samples. Indeed, all the low-$z$ external historical sample from previous surveys is in the first subset ($z<0.1$), while the supernovae measured by the full five years of DES are mostly at $z>0.1$ (with the exception of 3 supernovae). We refer to figure \ref{fig:composition} for the composition of DES5Y as inferred from data in \cite{DES-data}.

\bigskip

Ref.~\cite{Efstathiou:2024xcq} argues that the larger preference against $\Lambda$CDM, compared to Pantheon\texttt{+}, in the analysis of~\cite{DESI:2024mwx} is due to such an offset. Our analysis will be more general, as we will show in the next subsections, but it is interesting to inspect the presence of the offset again simply by studying the magnitudes $\mu$ of the DES5Y sample (for the SNe in common with Pantheon\texttt{+}). As a preliminary step therefore we compare $\mu^{\rm DES/PAN}$ with the magnitudes computed in a flat $\Lambda$CDM reference model (with $\Omega_m= 0.306$), defining $\Delta \mu \equiv \mu- \mu_{\rm \Lambda CDM}$. We proceed to binning $\Delta\mu$ (see appendix) and we report the binned sample in the left panel of figure \ref{fig:PAN_DES_common}, where  the offset is rather visible. Also, by applying a weighted average $\langle ... \rangle $ to the binned $\Delta \mu$ in the whole range of redshifts, we find that in the DES5Y sample in common with Pantheon\texttt{+}, $\langle \Delta \mu^{\rm DES} \rangle_{z<0.1}-\langle \Delta \mu^{\rm DES} \rangle_{z>0.1}=0.045$. Here the weighted average is done with errors taken from the diagonal of the binned covariance matrix. Moreover, we note that  in DES5Y all such common supernovae have smaller error bars than for Pantheon\texttt{+}, by an average overall factor of about 1.15.\footnote{Such error bars are the ones that come from the diagonal of the covariance matrix in Pantheon\texttt{+}. Note however that this does not match with the error bars on the distance moduli provided by the collaboration \cite{PAN-data}, which are even larger, on average by another factor of 1.4.} This is most likely due to different light-curve fitting models and, in particular, to the use of different wavelength ranges.~\footnote{M.Vincenzi, private communication.} In Figure~\ref{fig:PAN_DES_common}, we also observe that the correlation matrix (binned) reveals a significantly different pattern of correlations between the DES5Y and Pantheon datasets (in common). Notably, strong correlations are found among the very low-z sample in the DES5Y dataset, as seen in the center panel of Figure~\ref{fig:PAN_DES_common}.

\medskip

Finally, before going to the quantitative  analysis, we wish to comment on the full DES5Y catalog displayed in figure \ref{fig:composition}. Also in the full sample, the nature of the low/high-$z$ samples is rather different. We again proceed in defining $\Delta\mu$ and binning the full sample with the usual weights. From the correlation matrix of the full sample (see figure \ref{fig:composition}) we again see the strong correlation at low-$z$, and sizable correlation also in the very high-$z$ bins, $z\gtrsim 0.5-0.6$, of the sample. This suggests to split the DES5Y catalog in two samples, also exploiting the not so large correlation between them. 

\bigskip
\paragraph{A new split likelihood for DES5Y}~\\
Taking inspiration from the discussion so far, we consider the possibility of splitting the DES5Y dataset in two subsets.
We introduce a relative offset $\delta M$ between low and high $z$ SN (we call the resulting modified dataset DES5Y$_{\rm split}$), 
\be
\text{DES5Y}_{\text{split}} = \text{DES5Y}_{z<0.1, M + \delta M} \oplus \text{DES5Y}_{z>0.1, M}\,.
\ee
In practice, we have added a nuisance parameter $\delta M$ to the magnitude of the low-$z$ population.

\begin{figure}[h!]
  \centering
  \includegraphics[width=0.74\textwidth]{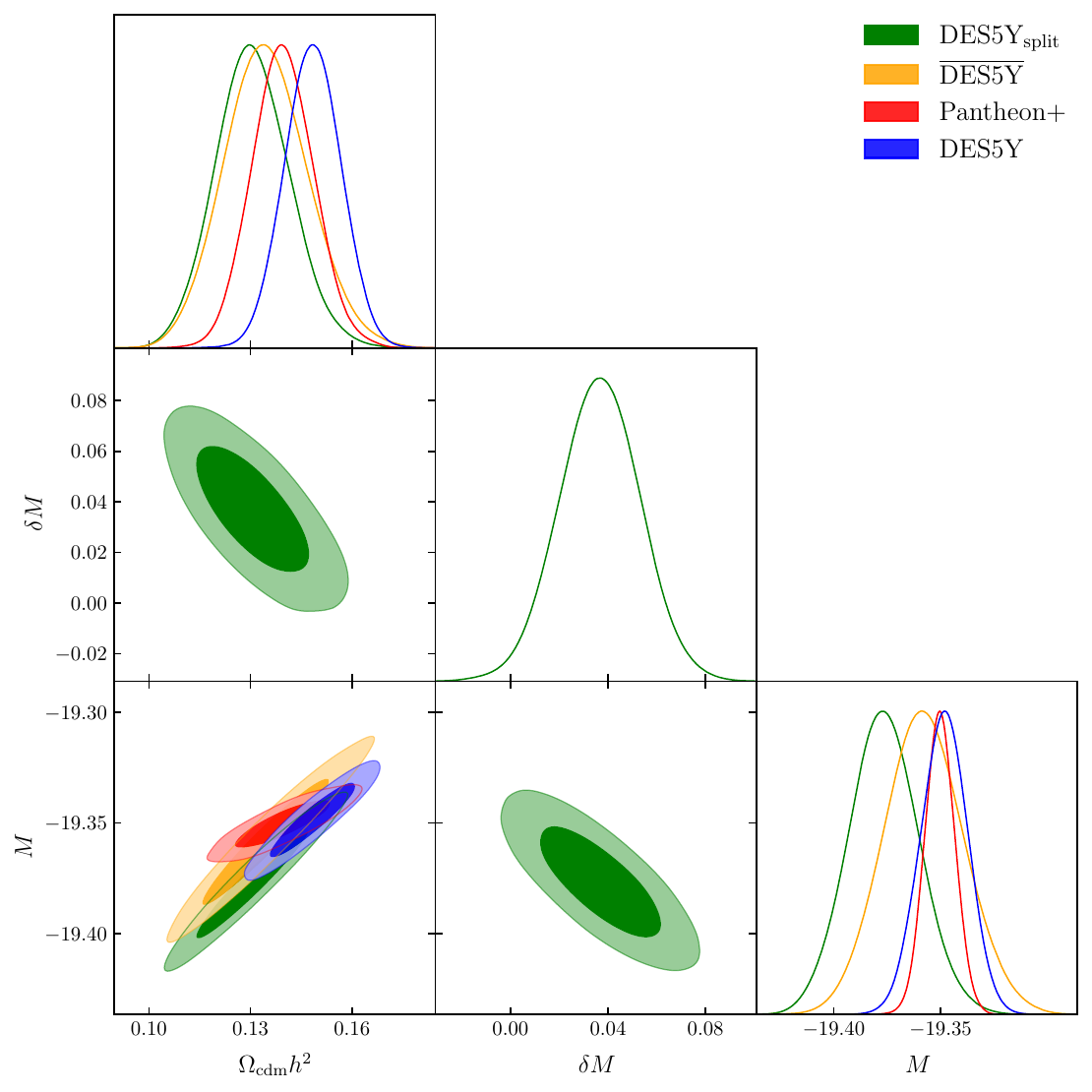}
\caption{\it \small Marginalized posterior distributions for the parameters $\Omega_{\rm cdm}$, $M$ and $\delta M$ in $\Lambda$CDM model. All input parameters have uniform priors. We have only included in the fit the SNe datasets, corresponding to  included are P18 + DESI-BAO and a Supernovae sample. We have tested four choices of SNe samples: DES5Y in blue, Pantheon\texttt{+} in red,  DES5Y$_{\text{split}}$ in green, and $\overline{\text{DES5Y}}$ in orange. Cfr. tables \ref{tab:all-data} and \ref{tab:mytable} and sections \ref{sec:COMBO} and \ref{sec:SPLIT}. We have redefined the scale of $M$ in $\overline{\text{DES5Y}}$ by subtracting 19.37, in order to plot it together with Pantheon\texttt{+}.} 
        \label{fig:DES_split_errors_LCDM}
\end{figure}

Allowing $\delta M$ to vary we first show in a simple $\Lambda$CDM fit with DES5Y$_{\rm split}$ alone that $\Omega_m$ gets smaller, and actually even smaller than the Pantheon\texttt{+} value, see figure \ref{fig:DES_split_errors_LCDM}.  Note also that a fit with $\overline{\text{DES5Y}}$ alone yields a result for $\Omega_m$ in the same direction, which is consistent. This is because marginalizing over $\delta M$ effectively ignores the overall low-$z$ mean magnitude, a procedure similar to discarding the common sample, which contains almost all of the low-$z$ data.

We then compare this new dataset with the usual DES5Y, fitting $\Lambda$CDM and $w_0 w_a$CDM against the following full data combinations:
\be\label{split}
\begin{split}
&1)\quad \mathrm{P18+DESI+DES5Y}\,,\\
&2)\quad \mathrm{P18+DESI+DES5Y_{split}}\,,\\
\end{split}
\ee

\begin{figure}[t!]
  \centering
  \includegraphics[width=0.8\textwidth]{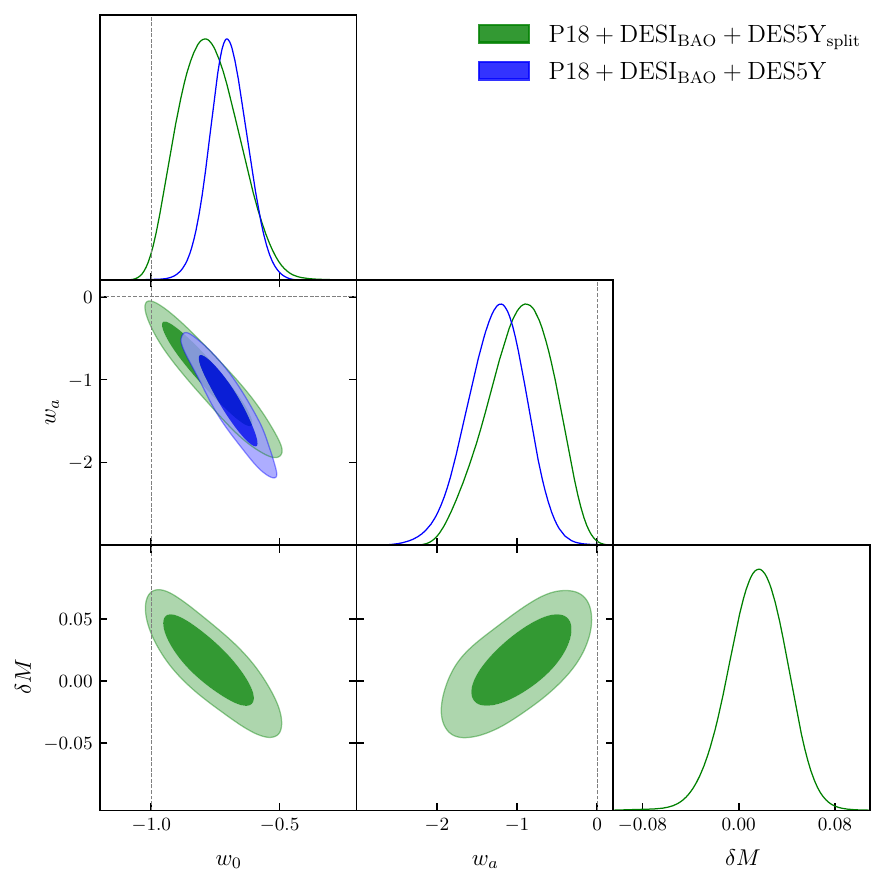}
\caption{\it Marginalized posterior distributions for the parameters $w_0$, $w_a$ and $H_0$ in the $w_aw_0$CDM model. All input parameters have uniform priors. The datasets included are P18 + DESI-BAO and a Supernovae sample. In particular we have tested the DES5Y$_{\text{split}}$ sample (in green) compared to DES5Y (in blue). Their definitions can be found tables \ref{tab:all-data} and \ref{tab:mytable} as well as in section \ref{sec:SPLIT}.} 
\label{fig:DES_split}
\end{figure}

In fig.~\ref{fig:DES_split} we show that indeed in the $w_0w_a$CDM model  a nonzero shift is preferred, with $\delta M=0.037\pm0.017$, and $\Lambda$CDM becomes now disfavored  at $1.7~\sigma$ only.

\newpage

\section{Conclusions}\label{sec:conclusions}
At present, the sensitivity to  time variations of dark energy relies on measurements of the BAO peak and spectrum as well as on (luminosity) distances inferred from supernovae observations. In this work, we re-examined the evidence disfavouring the $\Lambda$CDM model, showing the impact of slightly different treatments of the datasets. Our conclusions can be summarized in the following table:
   
\begin{center}
\renewcommand{\arraystretch}{1.5}  
\setlength{\tabcolsep}{10pt}    

\begin{tabular}{>{\columncolor[gray]{0.9}}l|>{\columncolor[gray]{0.97}}c|>{\columncolor[gray]{0.97}}c}
\hline
\rowcolor[HTML]{3D3E50}
\textcolor{white}{\textbf{Dataset}} & \textcolor{white}{\textbf{$\chi^2_{\rm min}(w_0w_a\mathrm{CDM})$}} & \textcolor{white}{\textbf{$\Lambda$CDM} exclusion}  \\ \hline

\small P18+DESI$_{\rm BAO}$+${\rm DES5Y}$  & 4431 & 3.9$\sigma$  \\ 

\small P18+DESI$_{\rm BAO}$+${\rm Pantheon\texttt{+}}$ & 4205 & 2.5$\sigma$ \\ 

\small P18+DESI$_{\rm BAO}$+$\overline{{\rm DES5Y}} + {\rm Pantheon\texttt{+}}$   & 5550 & 2.5$\sigma$  \\ 

\small P18+DESI$_{\rm BAO}$+$\overline{{\rm Pantheon\texttt{+}}} + {\rm DES5Y}$   & 5569 & 3.8$\sigma$  \\ \hline

\small P18+DESI$_{\rm BAO}$+${\rm DES5Y}_{\rm split}$   & 4431  & 1.7$\sigma$ \\ \hline

\small P18+DES$_{\rm BAO}$+${\rm DES5Y}$  & 4419   & 3$\sigma$ \\ 
\end{tabular}
\end{center}

First, we have shown that the evidence in favour of a $w_0w_a$CDM parametrization of dark energy can be determined independently of the DESI$_{\rm BAO}$ sample. To support this 
we replaced the DESI$_{\rm BAO}$ data with the single BAO data point at $z=0.85$ measured by DES (combined with DES5Y supernovae and P18 CMB data) finding a 3$\sigma$ exclusion for $\Lambda$CDM.
This result provides additional independent support for evolving dark energy. 

Second, we have considered the possibility of combining different SN catalogs to increase the statistical significance. The combination is non-trivial due to the different treatment of the common SNe contained in Pantheon\texttt{+} and DES5Y catalogues. In particular, for DES5Y we have shown that the treatment of such common supernovae is crucial: if we remove them from DESY5 and then combine with Pantheon\texttt{+}, we find overall agreement with results obtained using Pantheon\texttt{+} alone, such  that $\Lambda$CDM is only mildly  disfavoured (at $2.5\sigma$, in combination with P18+DESI$_{\rm BAO}$). If instead we perform the opposite combination, i.e. remove the common SNe from Pantheon\texttt{+} and combining with the full DESY5 dataset, we find that $\Lambda$CDM is excluded at 3.8$\sigma$ (in combination with P18+DESI$_{\rm BAO}$), similar to the fit without Pantheon\texttt{+}.

The different result of the SNe combination raises the question of the consistency of the catalogs. We have shown -- campatibly with the observation of Ref.~\cite{Efstathiou:2024xcq} -- that the disagreement can be accounted for by allowing different offsets in the DES5Y dataset for supernovae at redshifts $z>0.1$ (where most of the new measurements are) and $z<0.1$ (where all the old measurements are). Once this adjustment is made, the evidence against $\Lambda$CDM weakens to $1.7\sigma$.

To conclude, while the BAO measurement by DES lends support to evolving dark energy, we believe that the consistency of the supernova sample from DES 
needs to be fully investigated in order to draw firm conclusion disfavouring $\Lambda$CDM based on combinations that include such a dataset.

\subsubsection*{Acknowledgements}
We wish to thank Elisabeta Lusso, Guido Risaliti and Marko Simonovi\'c for interesting discussions and Maria Vincenzi for feedback on the DES5Y dataset.
We acknowledge the use of the computing resources provided by the ``PC-Farm'' at INFN Florence and by the ``Nyx" cluster at ICCUB, Barcelona. 

\appendix

\section{Details of MCMC runs}\label{app:tables}
Here we list the outcome of all our MCMCs for the $w_0w_a$CDM parametrization ($w_0w_a$CDM model), for all the data combinations. 
In all our chains we assume uniform priors on the input parameters. We also report the $\Delta \chi^2$ with respect to $\Lambda$CDM for the
same dataset.

\begin{table}[h]
    \centering
    \caption{\textbf{$\mathbf{w_0w_a}$CDM model with P18+DES5Y+DESI$_{\text{BAO}}$ datasets.}}   
        \label{tab:w0waDESBAO}
    \renewcommand{\arraystretch}{1.1}  
    \setlength{\tabcolsep}{10pt}      
        \begin{tabular}{>{\columncolor[gray]{0.9}}l|c|c|c|c}
        \hline
        \rowcolor[HTML]{3D3E50}
        \textcolor{white}{\textbf{Parameter}} & 
        \textcolor{white}{\textbf{Best-fit}} & 
        \textcolor{white}{\textbf{Mean $\pm \sigma$}} & 
        \textcolor{white}{\textbf{95\% Lower}} & 
        \textcolor{white}{\textbf{95\% Upper}} \\ 
        \hline
        $100~\omega_{b}$        & $2.234$   & $2.237_{-0.014}^{+0.013}$ & $2.209$ & $2.264$ \\
      $\omega_{\rm cdm }$         & $0.1202$  & $0.1198_{-0.00097}^{+0.0011}$ & $0.1178$ & $0.1218$ \\
        $100\theta_{s}$         & $1.042$   & $1.042_{-0.00029}^{+0.00029}$ & $1.041$ & $1.042$ \\
$\ln(10^{10}A_{s})$     & $3.052$   & $3.045_{-0.015}^{+0.015}$ & $3.016$ & $3.076$ \\
        $n_{s}$                 & $0.966$   & $0.965_{-0.004}^{+0.0036}$ & $0.9575$ & $0.9726$ \\
        $\tau_{\rm reio}$           & $0.05776$ & $0.05494_{-0.0079}^{+0.0076}$ & $0.03956$ & $0.07014$ \\ \hline
        $w_0$                   & $-0.7259$ & $-0.6948_{-0.076}^{+0.073}$ & $-0.8402$ & $-0.5494$ \\
        $w_a$                   & $-1.249$  & $-1.275_{-0.32}^{+0.4}$ & $-1.981$ & $-0.5855$ \\
           $\Omega_{\rm DE}$ &$0.6862$ & $0.6835_{-0.0064}^{+0.0073}$ & $0.6702$ & $0.6974$ \\ 
        \hline
        $M_{\rm DES}$                     & $-0.04$   & $-0.04182_{-0.017}^{+0.017}$ & $-0.07421$ & $-0.008567$ \\
       $H_0 \mathrm{[km/s/Mpc]}$                 & $67.73$   & $67.33_{-0.64}^{+0.69}$ & $66.03$ & $68.67$ \\
        $\sigma_8$              & $0.8158$  & $0.808_{-0.01}^{+0.012}$ & $0.7846$ & $0.8304$ \\
    \end{tabular} \\
    \vspace{0.5em}
    $-\ln{\cal L}_\mathrm{min} = 2215.31$, minimum $\chi^2 = 4431$,  $\Delta\chi^2=-18$
\end{table}

\begin{table}[h]
    \centering
    \caption{ \textbf{$\mathbf{w_0w_a}$CDM model with P18+DES5Y+DES$_{\text{BAO}}$ datasets.}}   
        \label{tab:w0waDESBAO}
    \renewcommand{\arraystretch}{1.05}  
    \setlength{\tabcolsep}{10pt}  
        \begin{tabular}{>{\columncolor[gray]{0.9}}l|c|c|c|c}
        \hline
        \rowcolor[HTML]{3D3E50}
        \textcolor{white}{\textbf{Parameter}} & 
        \textcolor{white}{\textbf{Best-fit}} & 
        \textcolor{white}{\textbf{Mean $\pm \sigma$}} & 
        \textcolor{white}{\textbf{95\% Lower}} & 
        \textcolor{white}{\textbf{95\% Upper}} \\ 
        $100\,\omega_b$ & $2.237$   & $2.238_{-0.016}^{+0.015}$    & $2.208$   & $2.268$   \\
      $\omega_{\rm cdm }$ & $0.1197$  & $0.1199_{-0.0013}^{+0.0012}$ & $0.1175$  & $0.1223$  \\
        $100\theta_s$   & $1.042$   & $1.042_{-0.00029}^{+0.0003}$ & $1.041$   & $1.042$   \\
$\ln(10^{10}A_{s})$  & $3.046$ & $3.046_{-0.015}^{+0.015}$    & $3.016$   & $3.076$   \\
        $n_s$           & $0.9673$  & $0.9655_{-0.0043}^{+0.0043}$ & $0.9572$  & $0.9739$  \\
        $\tau_{reio}$   & $0.05378$ & $0.05482_{-0.0079}^{+0.0073}$ & $0.03961$ & $0.07025$ \\
            \hline
        $w_0$           & $-0.6401$ & $-0.6648_{-0.096}^{+0.096}$ & $-0.8529$ & $-0.4745$ \\
        $w_a$           & $-1.393$  & $-1.483_{-0.47}^{+0.52}$    & $-2.465$  & $-0.4924$ \\
   \hline
        $M$             & $-0.03807$ & $-0.03245_{-0.027}^{+0.03}$ & $-0.0886$ & $0.02331$ \\
        $H_0 \mathrm{[km/s/Mpc]}$           & $67$      & $67.6_{-0.91}^{+0.93}$       & $65.81$   & $69.42$   \\
        $\sigma_8$      & $0.8104$  & $0.8098_{-0.012}^{+0.015}$   & $0.7822$  & $0.8358$  \\
        
    \end{tabular} \\
    \vspace{0.5em}
    $-\ln{\cal L}_\mathrm{min} = 2209.71$, minimum $\chi^2 = 4419$, $\Delta\chi^2 = -12$
\end{table}

\newpage

\begin{table}[h]
    \centering
    \caption{ \textbf{$\mathbf{w_0w_a}$CDM model with P18+DES5Y+BOSS$_{\text{BAO}}$ datasets.}}   
     \label{tab:w0waDESBAO}
    \renewcommand{\arraystretch}{1.05}  
    \setlength{\tabcolsep}{10pt}    
    \begin{tabular}{>{\columncolor[gray]{0.9}}l|c|c|c|c}
        \hline
        \rowcolor[HTML]{3D3E50}
        \textcolor{white}{\textbf{Parameter}} & 
        \textcolor{white}{\textbf{Best-fit}} & 
        \textcolor{white}{\textbf{Mean $\pm \sigma$}} & 
        \textcolor{white}{\textbf{95\% Lower}} & 
        \textcolor{white}{\textbf{95\% Upper}} \\ 
        \hline
$100~\omega_b$ &$2.239$ & $2.232_{-0.015}^{+0.014}$ & $2.203$ & $2.26$ \\ 
$\omega_{\rm cdm}$ &$0.1208$ & $0.1206_{-0.0011}^{+0.0011}$ & $0.1184$ & $0.1228$ \\ 
$100\theta_{s }$ &$1.042$ & $1.042_{-0.0003}^{+0.0003}$ & $1.041$ & $1.042$ \\ 
$\ln(10^{10}A_{s})$  &$3.045$ & $3.047_{-0.016}^{+0.015}$ & $3.017$ & $3.077$ \\ 
$n_{s }$ &$0.9617$ & $0.9632_{-0.004}^{+0.004}$ & $0.9552$ & $0.971$ \\ 
$\tau_{\rm reio }$ &$0.05274$ & $0.05467_{-0.008}^{+0.0074}$ & $0.03939$ & $0.07022$ \\ \hline
$w_0$ &$-0.8096$ & $-0.7687_{-0.081}^{+0.071}$ & $-0.9204$ & $-0.6148$ \\ 
$w_a$ &$-0.609$ & $-0.948_{-0.3}^{+0.45}$ & $-1.763$ & $-0.2027$ \\ 
$\Omega_{\rm DE }$ &$0.671$ & $0.6735_{-0.0069}^{+0.0079}$ & $0.6587$ & $0.688$ \\ 
 \hline
$M$ &$-0.08641$ & $-0.07236_{-0.016}^{+0.016}$ & $-0.1036$ & $-0.04093$ \\ 
       $H_0 \mathrm{[km/s/Mpc]}$  &$66.16$ & $66.53_{-0.63}^{+0.62}$ & $65.3$ & $67.77$ \\ 
$\sigma_8$ &$0.8059$ & $0.7998_{-0.011}^{+0.014}$ & $0.7741$ & $0.8239$ \\
    \end{tabular} \\
    \vspace{0.5em}
$-\ln{\cal L}_\mathrm{min} =2213.49$, minimum $\chi^2=4427$
\end{table}

\begin{table}[h]
    \centering
    \caption{ \textbf{$\mathbf{w_0w_a}$CDM model with P18+Pantheon\texttt{+}+DESI$_{\text{BAO}}$datasets.}}   
     \label{tab:w0waDESBAO}
    \renewcommand{\arraystretch}{1.05}  
        \setlength{\tabcolsep}{10pt}       
    \begin{tabular}{>{\columncolor[gray]{0.9}}l|c|c|c|c}
        \hline
        \rowcolor[HTML]{3D3E50}
        \textcolor{white}{\textbf{Parameter}} & 
        \textcolor{white}{\textbf{Best-fit}} & 
        \textcolor{white}{\textbf{Mean $\pm \sigma$}} & 
        \textcolor{white}{\textbf{95\% Lower}} & 
        \textcolor{white}{\textbf{95\% Upper}} \\ 
        \hline
   $100~\omega_{b }$ &$2.237$ & $2.238_{-0.014}^{+0.014}$ & $2.211$ & $2.266$ \\ 
$\omega{}_{cdm }$ &$0.12$ & $0.1196_{-0.0011}^{+0.00096}$ & $0.1178$ & $0.1216$ \\ 
$100\theta_{s }$ &$1.042$ & $1.042_{-0.00028}^{+0.00029}$ & $1.041$ & $1.042$ \\ 
$\ln(10^{10}A_{s})$  &$3.028$ & $3.043_{-0.015}^{+0.014}$ & $3.016$ & $3.073$ \\ 
$n_{s }$ &$0.964$ & $0.9656_{-0.0036}^{+0.004}$ & $0.9581$ & $0.9729$ \\ 
$\tau_{\rm reio }$ &$0.04555$ & $0.05413_{-0.0075}^{+0.0065}$ & $-$ & $-$ \\ \hline
$w_0$ &$-0.7994$ & $-0.8128_{-0.067}^{+0.062}$ & $-0.9378$ & $-0.6849$ \\ 
$w_a$ &$-0.8319$ & $-0.8325_{-0.24}^{+0.32}$ & $-1.401$ & $-0.2897$ \\ 
$\Omega_{\rm DE }$ &$0.6895$ & $0.694_{-0.0066}^{+0.0064}$ & $0.6817$ & $0.7065$ \\ 
\hline
$M$ &$-19.4$ & $-19.39_{-0.018}^{+0.019}$ & $-$ & $-$ \\ 
        $H_0 \mathrm{[km/s/Mpc]}$ &$67.88$ & $68.3_{-0.68}^{+0.68}$ & $67$ & $69.6$ \\ 
$\sigma_8$ &$0.8132$ & $0.8216_{-0.01}^{+0.0095}$ & $0.8016$ & $0.8417$ \\       
    \end{tabular} \\
    \vspace{0.5em}
 $-\ln{\cal L}_\mathrm{min} =2102.41$, minimum $\chi^2=4205$,  $\Delta\chi^2=-8.7$
\end{table}

\newpage

\begin{table}[h]
    \centering
    \caption{ \textbf{$\mathbf{w_0w_a}$CDM model with P18+Pantheon\texttt{+}\,+$\overline{\text{DES5Y}}$+DESI$_{\text{BAO}}$ datasets.}}   
      \label{tab:w0waDESBAO}
    \renewcommand{\arraystretch}{1.05}  
    \setlength{\tabcolsep}{10pt}     
    \begin{tabular}{>{\columncolor[gray]{0.9}}l|c|c|c|c}
        \hline
        \rowcolor[HTML]{3D3E50}
        \textcolor{white}{\textbf{Parameter}} & 
        \textcolor{white}{\textbf{Best-fit}} & 
        \textcolor{white}{\textbf{Mean $\pm \sigma$}} & 
        \textcolor{white}{\textbf{95\% Lower}} & 
        \textcolor{white}{\textbf{95\% Upper}} \\ 
        \hline
 $100~\omega{}_{b }$ &$2.24$ & $2.239_{-0.014}^{+0.014}$ & $2.212$ & $2.266$ \\ 
$\omega{}_{cdm }$ &$0.1194$ & $0.1195_{-0.001}^{+0.001}$ & $0.1175$ & $0.1215$ \\ 
$h$ &$0.6783$ & $0.6798_{-0.0065}^{+0.0063}$ & $0.6676$ & $0.6921$ \\ 
$\ln(10^{10}A_{s})$   &$3.052$ & $3.045_{-0.015}^{+0.014}$ & $3.016$ & $3.074$ \\ 
$n_{s }$ &$0.9658$ & $0.9658_{-0.004}^{+0.0038}$ & $0.9582$ & $0.9733$ \\ 
$\tau_{\rm reio }$ &$0.05922$ & $0.05489_{-0.0075}^{+0.007}$ & $0.04006$ & $0.06972$ \\ \hline
$w_0$ &$-0.8013$ & $-0.8308_{-0.062}^{+0.059}$ & $-0.9465$ & $-0.7123$ \\ 
$w_a$ &$-0.7969$ & $-0.7134_{-0.24}^{+0.28}$ & $-1.226$ & $-0.2132$ \\ 
$\Omega_{\rm DE }$ &$0.6903$ & $0.6914_{-0.0062}^{+0.0063}$ & $0.6792$ & $0.7034$ \\ 
\hline
$M$ &$-19.4$ & $-19.4_{-0.018}^{+0.018}$ & $-19.44$ & $-19.37$ \\ 
$\delta M$ &$0.004688$ & $-0.0008988_{-0.009}^{+0.0088}$ & $-0.01838$ & $0.01663$ \\ 
        $H_0 \mathrm{[km/s/Mpc]}$ &$67.83$ & $67.98_{-0.65}^{+0.63}$ & $66.76$ & $69.21$ \\        
    \end{tabular} \\
    \vspace{0.5em}
$-\ln{\cal L}_\mathrm{min} =2774.81$, minimum $\chi^2=5550$,  $\Delta\chi^2=-8.6$
\end{table}

\begin{table}[h]
    \centering
    \caption{ \textbf{$\mathbf{w_0w_a}$CDM model with P18+DES5Y+$\overline{\text{Pantheon\texttt{+} }}$+DESI$_{\text{BAO}}$datasets.}}   
     \label{tab:w0waDESBAO}
    \renewcommand{\arraystretch}{1.05} 
    \setlength{\tabcolsep}{10pt}    
    \begin{tabular}{>{\columncolor[gray]{0.9}}l|c|c|c|c}
        \hline
        \rowcolor[HTML]{3D3E50}
        \textcolor{white}{\textbf{Parameter}} & 
        \textcolor{white}{\textbf{Best-fit}} & 
        \textcolor{white}{\textbf{Mean $\pm \sigma$}} & 
        \textcolor{white}{\textbf{95\% Lower}} & 
        \textcolor{white}{\textbf{95\% Upper}} \\ 
        \hline
   $100~\omega_{b }$ &$2.231$ & $2.239_{-0.014}^{+0.013}$ & $-$ & $-$ \\ 
$\omega_{\rm cdm }$ &$0.12$ & $0.1196_{-0.00097}^{+0.00089}$ & $0.1177$ & $0.1215$ \\ 
$h$ &$0.6745$ & $0.6765_{-0.0052}^{+0.0063}$ & $0.6654$ & $0.6873$ \\ 
$\ln(10^{10}A_{s})$   &$3.039$ & $3.043_{-0.015}^{+0.013}$ & $3.015$ & $3.072$ \\ 
$n_{s }$ &$0.965$ & $0.9656_{-0.0035}^{+0.0037}$ & $0.9582$ & $0.9726$ \\ 
$\tau_{\rm reio }$ &$0.05334$ & $0.05444_{-0.0076}^{+0.0071}$ & $-$ & $-$ \\ \hline
$w_0$ &$-0.787$ & $-0.7727_{-0.05}^{+0.059}$ & $-0.8839$ & $-0.6712$ \\ 
$w_a$ &$-0.8384$ & $-0.9048_{-0.22}^{+0.27}$ & $-1.404$ & $-0.4223$ \\ 
$\Omega_{\rm DE }$ &$0.6857$ & $0.6882_{-0.0056}^{+0.0062}$ & $-$ & $-$ \\ 
\hline
$M$ &$-19.42$ & $-19.41_{-0.016}^{+0.016}$ & $-19.44$ & $-19.38$ \\ 
$\delta M$ &$0.0106$ & $0.00939_{-0.0069}^{+0.0075}$ & $-0.004954$ & $0.02307$ \\ 
        $H_0 \mathrm{[km/s/Mpc]}$  &$67.45$ & $67.65_{-0.52}^{+0.63}$ & $66.54$ & $68.73$ \\       
    \end{tabular} \\
    \vspace{0.5em}
$-\ln{\cal L}_\mathrm{min} =2784.73$, minimum $\chi^2=5569$,  $\Delta\chi^2=-17.4$
\end{table}

\newpage

\begin{table}[h]
    \centering
    \caption{ \textbf{$\mathbf{w_0w_a}$CDM model with P18+DES5Y$_{\rm split}$+DESI$_{\text{BAO}}$ datasets.}}   
    \label{tab:w0waDESBAO}
    \renewcommand{\arraystretch}{1.05}  
    \setlength{\tabcolsep}{10pt}    
    \begin{tabular}{>{\columncolor[gray]{0.9}}l|c|c|c|c}
        \hline
        \rowcolor[HTML]{3D3E50}
        \textcolor{white}{\textbf{Parameter}} & 
        \textcolor{white}{\textbf{Best-fit}} & 
        \textcolor{white}{\textbf{Mean $\pm \sigma$}} & 
        \textcolor{white}{\textbf{95\% Lower}} & 
        \textcolor{white}{\textbf{95\% Upper}} \\ 
        \hline
 $100~\omega_{b }$ &$2.238$ & $2.238_{-0.014}^{+0.014}$ & $2.21$ & $2.266$ \\ 
$\omega_{\rm cdm }$ &$0.1201$ & $0.1197_{-0.001}^{+0.0011}$ & $0.1177$ & $0.1217$ \\ 
$h$ &$0.682$ & $0.6798_{-0.01}^{+0.01}$ & $0.6594$ & $0.6999$ \\ 
$\ln(10^{10}A_{s })$ &$3.038$ & $3.043_{-0.015}^{+0.014}$ & $3.015$ & $3.072$ \\ 
$n_{s }$ &$0.9634$ & $0.9653_{-0.0038}^{+0.0039}$ & $0.9579$ & $0.973$ \\ 
$\tau_{\rm reio }$ &$0.05257$ & $0.05417_{-0.0076}^{+0.007}$ & $0.03976$ & $0.0692$ \\ \hline
$w_0$ &$-0.8149$ & $-0.7753_{-0.13}^{+0.11}$ & $-0.9928$ & $-0.5755$ \\ 
$w_a$ &$-0.8401$ & $-0.956_{-0.37}^{+0.46}$ & $-1.74$ & $-0.2272$ \\ 
$\Omega_{\rm DE}$ &$0.6921$ & $0.6908_{-0.0097}^{+0.01}$ & $0.6711$ & $0.7098$ \\ \hline
$M$ &$-0.04474$ & $-0.03981_{-0.016}^{+0.016}$ & $-0.07228$ & $-0.007477$ \\ 
$\delta M$ &$0.02411$ & $0.01547_{-0.023}^{+0.025}$ & $-0.03109$ & $0.06042$ \\ 
        $H_0 \mathrm{[km/s/Mpc]}$ &$68.2$ & $67.98_{-1}^{+1}$ & $65.94$ & $69.99$ \\ 
    \end{tabular} \\
    \vspace{0.5em}
$-\ln{\cal L}_\mathrm{min} =2215.59$, minimum $\chi^2=4431$,  $\Delta\chi^2=-4.9$
\end{table}

\section{Binned data samples}\label{app:binning}
In each SN catalog we have access to the magnitude values $\mu_i$ and their covariance matrix $\sigma_{ij}$ (including both statistical and systematical uncertainties), where the index runs over the full length of the catalog (see table \ref{tab:mytable}). In all the numerical analysis of our paper we have worked with unbinned data samples, to exploit fully the correlations among different variables. However, in our discussion, we often refer to binned data sample for visualization purposes. We use a weighted binning procedure to define new magnitude variables $\hat{\mu}_{\alpha}$, where now the greek index $\alpha$ runs over the bins
\be
\hat\mu_\alpha \equiv  \frac{\sum_i w_{\alpha,i} \mu_i}{\sum_k w_{\alpha,k}}\equiv \sum_i B_{\alpha,i}\mu_i\,,
\ee
where the rectangular matrix $B$ is such that $\sum_i B_{\alpha,i}=1$ for all the rows (bins) $\alpha$.
Here $w_{\alpha,i}$ are generic weights that go to zero if the $i$-th variable $\mu_i$ does not belong to the $\alpha$-th bin. Since our mapping is linear in the original variable, we can easily compute the covariance (and correlation) matrix $\hat\sigma_{\alpha\beta}$
of the new variables as
\be
\hat\sigma_{\alpha\beta}= \frac{\sum_i w_{\alpha,i}\sum_j w_{\beta, j} \sigma_{ij}}{(\sum_k w_{\alpha,k})(\sum_l w_{\beta,l})}\,, \quad \mathrm{corr}_{\alpha\beta}\equiv \frac{\hat\sigma_{\alpha\beta}}{\sqrt{\hat\sigma_{\alpha\alpha}\hat\sigma_{\beta\beta}}}\,.
\ee

\bigskip
In figures \ref{fig:PAN_DES_common} and \ref{fig:DES_split} we adopted a weighted binning so defined:
The edges of the bins are 
\be\label{eq:edges}\small
z_{\rm bin\ edges} = [0.02\,, 0.03\,, 0.04\,,0.05\,, 0.06\,, 0.07\,, 0.08\,, 0.09\,, 0.1\,, 0.2\,, 0.3\,, 0.4\,,0.5\,, 0.6\,, 0.7\,, 0.8\,, 0.9, 1\,, 1.1\,, 1.2] \,,
\ee
and the center of each bin, defined as the average between two consecutive entries in the above list,  $z_{\alpha}$, is used to plot the corresponding quantities.  The weights $w_{\alpha,i}$ are chosen to give more importance to the measurements $\mu_i$ with smaller errors, therefore we use the following expression
\be
w_{\alpha, i} = \sigma_{ii}^{-1}\, \quad \forall i \in \mathrm{bin}_\alpha\,.
\ee

\pagestyle{plain}
\bibliographystyle{jhep}
\small
\bibliography{biblio}

\end{document}